\documentclass[10pt,a4paper,twocolumn]{article}
\usepackage[utf8]{inputenc}
\usepackage{amsmath}
\usepackage{amsfonts}
\usepackage{amssymb}
\usepackage{graphicx}
\usepackage{authblk}
\usepackage{geometry}
\usepackage{color}
\usepackage{float}
\usepackage{miller}
\usepackage{gensymb}

\title{Low Temperature Ageing Behaviour of U-Nb $\alpha''$ Phase Alloys}
\author[1]{JE Sutcliffe}
\author[1]{CP Jones}
\author[1]{JE Darnbrough}
\author[1]{KR Hallam}
\author[1]{RS Springell}
\author[2]{P Ryan}
\author[2]{T Cartwright}
\author[1]{TB Scott}
\affil[1]{Interface Analysis Centre, HH Wills Physics Laboratory, University of Bristol, Tyndall Avenue, Bristol BS8 1TL, UK}
\affil[2]{AWE, Aldermaston, Reading, Berkshire RG7 4PR, UK}

\date{\today}
\begin{document}
 
\maketitle

\begin{abstract}
Ageing mechanisms of the U-5\,\%wtNb system have been investigated on samples exposed to temperatures of 150$\,\celsius$ for up to 5000\,hours. A variety of surface and bulk analytic techniques have been used to investigate phase, chemical and crystallographic changes. Characterisation of microstructural evolution was carried out through secondary electron microscopy (SEM), energy dispersive x-ray spectroscopy (EDS), electron backscatter diffraction (EBSD), transmission electron microscopy (TEM) and x-ray diffraction (XRD). This investigation suggests crystallographic defects such as twinning furthers the martensitic tendencies with ageing. Resizing of the lattice and shuffling of atoms results in a small progression from the $\alpha''$ towards the $\alpha'$ phase.
\end{abstract}

\section{Introduction}
Alloying uranium with niobium has been identified as a very successful method of improving a number of useful characteristics, such as greater corrosion resistance, irradiation resistance and ductility, a lower Young's modulus and the introduction of shape memory properties \cite{Wheeler2009,Jackson1970,Manner1999}. Besides the most common use in defence applications, $\gamma$-phase uranium alloys are being considered as potential low-enriched, high-density, fuels for use in research and test reactors \cite{Snelgrove1997,Meyer2002,Savchenko2010,Sinha2009,VandenBerghe2008,Clarke2015}.

Effective long-term storage and stockpiling of nuclear materials demands a strong understanding of the ageing behaviour and the resulting change in corrosion properties of uranium-based materials.

\begin{figure}
\centering
\includegraphics[width=\linewidth]{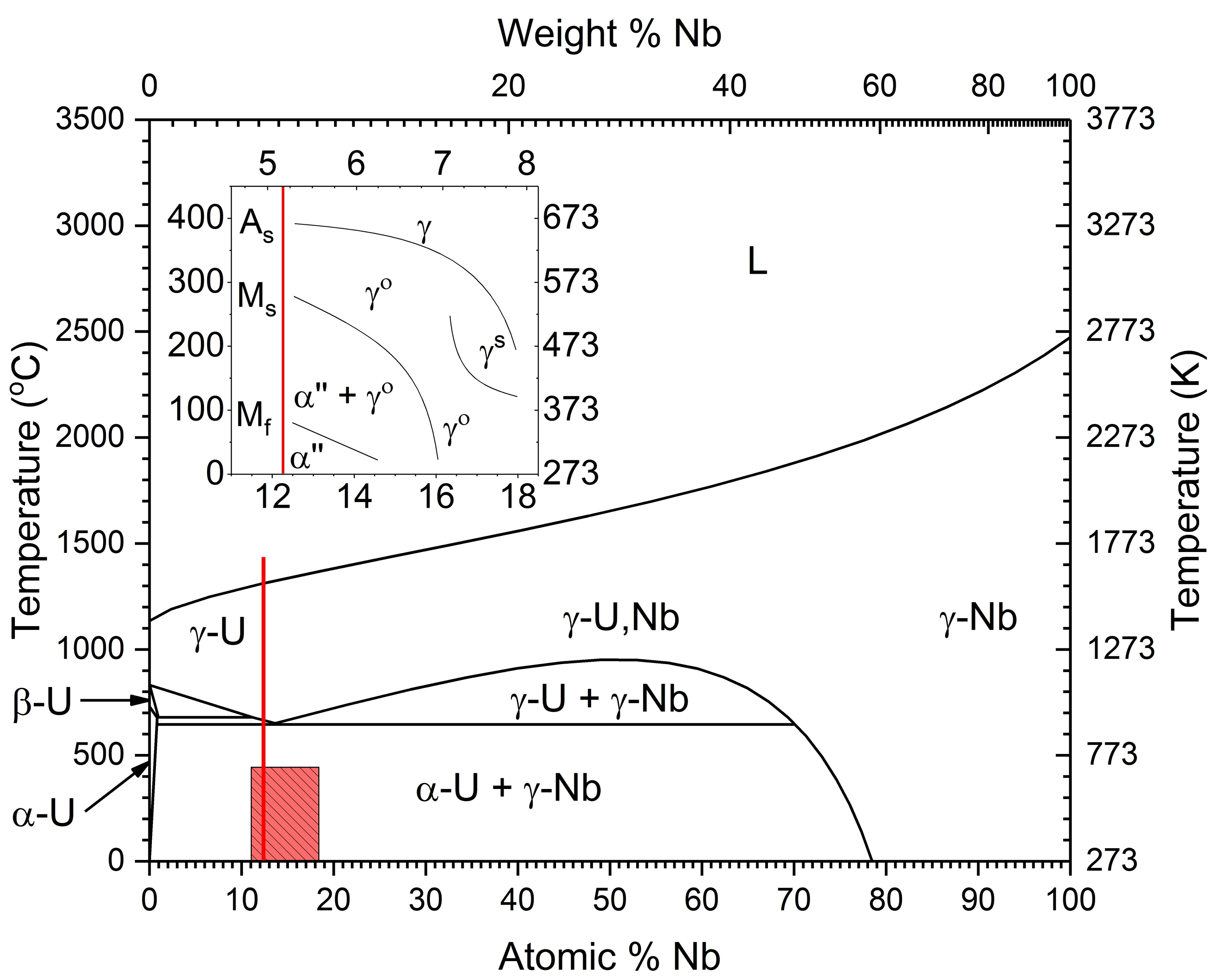}
\caption{Phase diagram of the U-Nb system. Top left inset shows the metastable states produced in the uranium-rich alloys produced by water quenching in the region denoted by the red box. Image constructed from the data of: \cite{Liu2008,Koike1998,Vandermeer1981}.}
\label{fig:Phase}
\end{figure}

Niobium is highly soluble with uranium in the high temperature $\gamma$ ($bcc$) phase, forming solid solutions above 950$\,\celsius$ \cite{Liu2008,Koike1998}. However, solubility in the $\alpha$ phase is limited to just 0.5\,wt\% \cite{Manner1999}, Figure \ref{fig:Phase}. Cooling below 647$\,\celsius$ exposes this metastability, causing phases to separate into an almost pure $\alpha$-uranium phase and a niobium rich $\gamma$-phase. This phase mixture exhibits poor corrosion resistance and less favourable mechanical properties \cite{Volz2007}.

Given sufficient alloying additions, due to slow diffusion of transition metals in uranium \cite{Fedorov1972}, retention of the $\gamma$ phase at room temperature may be achieved via a quench with a cooling rate of at least 20$\,\celsius$s$^{-1}$ \cite{Clarke2009}. For concentrations $\leq$ 6wt\%Nb, further cooling results in the $\gamma^{o}$ phase martensitically transforming to the monoclinic $\alpha''$ phase between 162 and 77$\,\celsius$, inset of Figure \ref{fig:Phase}. This martensitic structure features a high degree of twinning, thereby increasing ductility \cite{Wheeler2009}. The $\alpha''$ is metastable and decays when aged \cite{DAmato1964}. This work seeks to investigate the stability of the $\alpha''$ phase when exposed to accelerated ageing at moderately increased temperatures.

\section{Experimental Techniques}

\subsection{Materials}
Starting material supplied by AWE were cast from melt with 5.1\% niobium by weight, henceforth referred to as UNb5. Carbon is the most significant impurity, 153$\pm$28 ppm, but also oxygen and nitrogen, 42$\pm$4 and $<$4 ppm respectively. Following casting, a heat treatment was applied whereby the alloy was hot rolled to approximately 50\% of the original thickness. A further vacuum homogenisation treatment was conducted at 1000$\,\celsius$ for 6 hours followed by a 850$\,\celsius$ cooling step and water quench.

Unaged samples were compared to aged samples held at 150\,$\celsius$ for 1500, 3624 and 5000 hours. Stress-strain curves were obtained from dog-bone style tensile test specimens with an additional sample aged to 21600 hours. Unaged samples had been stored at ambient temperatures for approximately 6 years but had not undergone elevated temperature ageing.

All samples were initially prepared for investigation via a two step procedure of mechanical and electro-polishing as outlined in Jones \textit{et al.} \cite{Jones2015}. Argon ion etching was performed at 45$^{o}$ under UHV conditions to finish the preparation process. Etching at 4\,kV with a beam current of 4.2\,$\mu$A was found to be effective in removing any oxide whilst preserving the material underneath.

\subsection{Analytical Techniques}
X-ray diffraction was performed on a Philips X'Pert Pro MPD using Cu-K$\alpha$ radiation. The x-ray tube was operated at 40\,kV and 30\,mA. Specular $\theta$/2$\theta$ scans were performed between 20\,$^{o}$ and 50\,$^{o}$ with a step size of 0.02\,$^{o}$. CeO$_{2}$ powder was applied to the sample to calibrate the diffraction spectra.

Electron microscopy and EBSD were performed using a Zeiss EVO MA10 SEM fitted with a LaB$_{6}$ electron source, a Digiview 3 high speed camera and EDAX EBSD instrumentation. Orientation maps of up to 1\,mm$^{2}$ were obtained with step sizes of less than 2\,$\mu$m. The $\alpha''$ phase as given by Tangri and Chaudhuri \cite{Tangri1965}, was used as a reference phase to index against. Neighbour orientation correlation functions using OIM$^{TM}$ software were implemented when necessary.

TEM sections were extracted via FIB liftout using a Helios Nanolab 600i Dualbeam (DB). This is a two stage process utilising a high beam current to produce an initial lamella of material which is lifted out and secured to a Cu grid through ion-assisted platinum deposition before continuing thinning down to 50-100\,nm at progressively lower accelerating voltages. TEM lamellae were generated from unaged and 5000\,hr aged alloy samples.

DB milled slices were transferred to a Philips EM430 TEM with air exposure limited to less than 10 minutes. Images were taken predominately in bright field mode using an accelerating voltage of 250\,kV with diffraction patterns taking place using a 20$\mu$m selected area aperture and parallel (defocused) electron beam.

Hardness measurements were performed on polished samples using a Vickers microhardness tester fitted with a 200\,g load.

\section{Results \& Discussion} 

\subsection{Phase Identification}
\subsubsection{$\alpha$ Phases}
\begin{figure}
\centering
\includegraphics[width=0.5\textwidth]{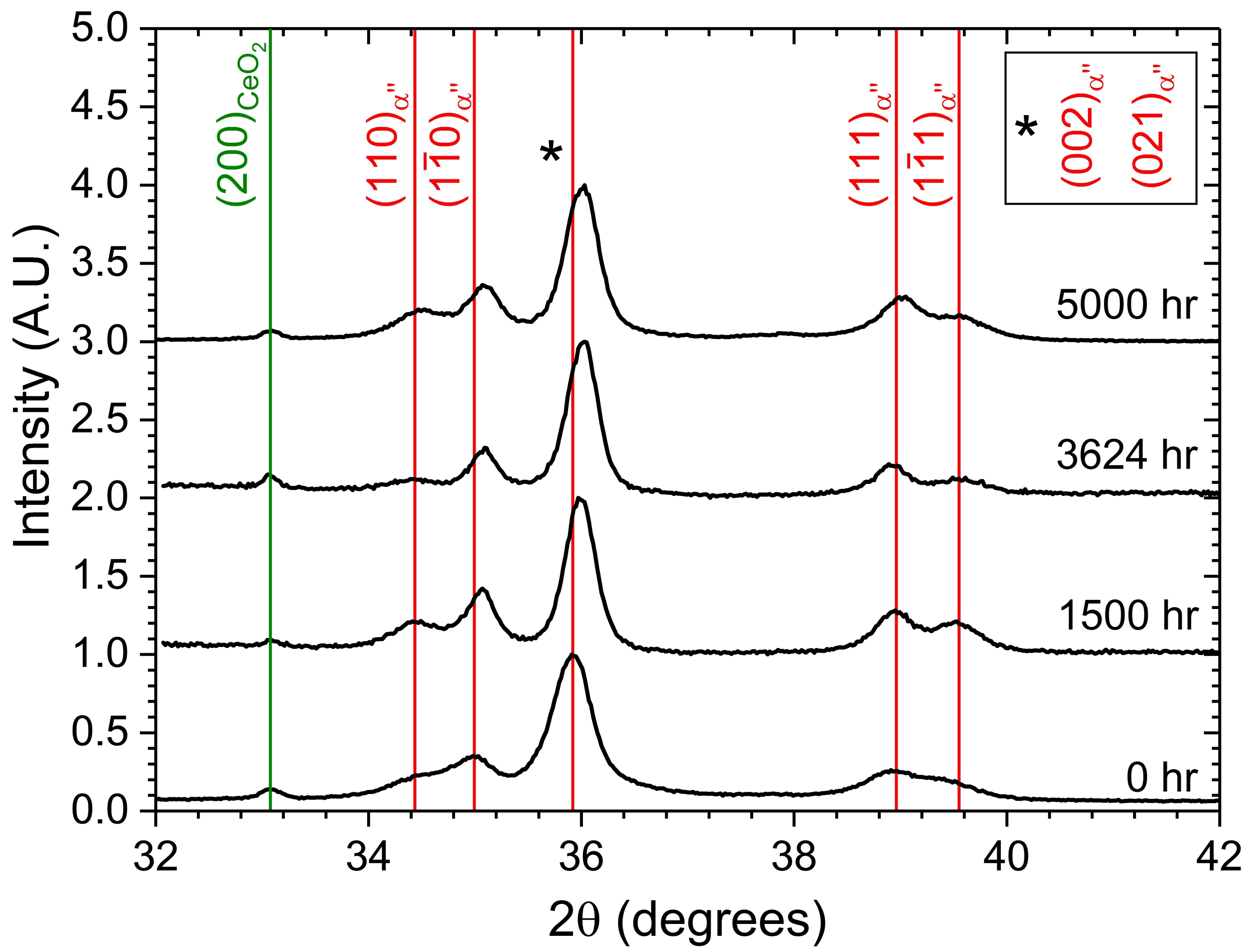}
\caption{X-ray diffraction patterns of samples aged for varying durations at 150$^{o}$C. Vertical lines represent 2$\theta$ positions corresponding to reflections from the unaged sample.} \label{fig:XRD}
\end{figure}

XRD measurements clearly demonstrate that the $\alpha''$ phase is the only observed structure in this study, though small changes in the lattice are experienced with ageing, Figure \ref{fig:XRD}. Specifically, a splitting of the \hkl(1-10)$_{\alpha''}$ and \hkl(110)$_{\alpha''}$ reflections and overall peak positions can be seen to slightly increase to larger values of 2$\theta$ suggesting that lattice strain is being induced.

The $\alpha''$ phase has been well characterised as possessing a monoclinic unit cell \cite{DAmato1964,Tangri1965,Anagnostidis1964}. Due to the non-conventional cell definition, this results in the characteristic splitting of various diffraction peaks concerning reflections involving both the $a$ and $b$ axes. Examples include the \hkl(1-10)$_{\alpha''}$-\hkl(110)$_{\alpha''}$ and \hkl(1-11)$_{\alpha''}$-\hkl(111)$_{\alpha''}$ peaks, Figure \ref{fig:XRD}. A separation of these peaks might indicate a very small increase in the $\gamma$-angle with ageing.

\begin{figure}
\centering
\includegraphics[width=0.5\textwidth]{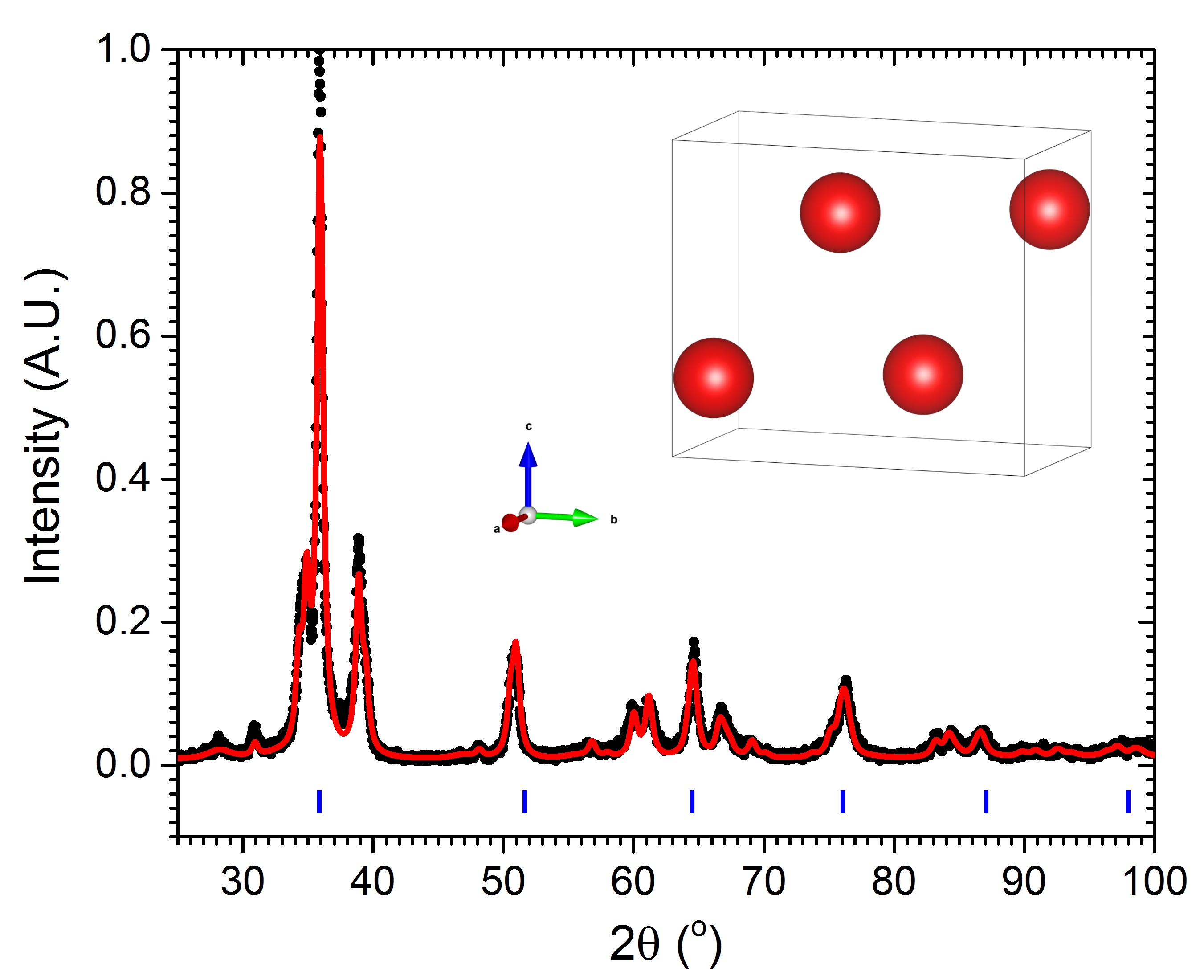}
\caption{X-ray diffraction pattern of the unaged sample. The pattern has been fitted with the $\alpha''$ phase in the C\,1\,1\,2$_{1}$/m space group as reported by Brown \textit{et al.} \cite{Brown2016}. Crucially, the \hkl(200)$_{\gamma}$ peak is not represented; \hkl(112)$_{\alpha''}$ and \hkl(1-12)$_{\alpha''}$ peaks feature at slightly larger $d$ spacings. Peaks positions of the $\gamma$ phase are shown by blue ticks below the pattern. Peak intensities suggest that there may be preferred orientation along certain directions.} \label{fig:XRD_2}
\end{figure}

Peak widths are observed to slightly decrease between the unaged and 1500 hr aged samples, Figure \ref{fig:XRD}. Peaks broadening might have been expected as a result of distortions of the $\alpha''$ structure in response to thermallyinduced stresses. Instead, though changes are minimal, it is possible that thermal treatment has undone stresses induced in quenching or ambient ageing, effectively annealing the material. Thermal treatment is unlikely to have exceeded the austenitic start temperature, A$_{s}$, Figure \ref{fig:Phase}, required to provide sufficient energy to undo the reversible displacive reactions.

A contraction of the lattice may be explained by a transition towards the $\alpha'$ and $\alpha$-U phase boundaries. Sliding down the metastable phase spectrum subtly affects unit cell lengths and the $\gamma$-angle. In this case, the $\gamma$-angle and $a$ and $c$ axes decrease while the $b$ axis increases resulting in a slight overall reduction in cell volume or densification \cite{Jackson1970}. It has been shown that the unit cell size decreases close to the $\alpha'$ boundary \cite{Jackson1970,Tangri1965,Anagnostidis1964}. A collective, uniform transition towards the $\alpha'$ phase may also be responsible for reducing variation in structure, reducing peak width. 

Overall, unit cell lattice parameters have changed rather insignificantly, featuring only small shifts to smaller $d$ spacings and narrowing of peaks. Since shifts occurred after the shortest ageing period, the role of instantaneous heating of the material must be thoroughly assessed.

\subsubsection{$\gamma$ Phases}

From Figure \ref{fig:XRD}, no new peaks that may be ascribed to either a significantly niobium rich $\gamma$, or niobium deficient $\alpha$-U phase emerge as a result of ageing. Additionally, there are no indications from the acquired XRD data to suggest that the metastable $\gamma^{o}$ phase has formed. The $\gamma$ to $\gamma^{o}$ reaction is known to split the \hkl(110)$_{\gamma}$ peak into the \hkl(220)$_{\gamma^{o}}$ and \hkl(201)$_{\gamma^{o}}$ reflections \cite{Volz2007}. Had the tetragonal $\gamma^{o}$ phase formed, a transformation of the \hkl(002)$_{\alpha''}$ and \hkl(021)$_{\alpha''}$ peaks into two comparatively sized peaks would be observed.

A full scan of the unaged alloy was conducted, Figure \ref{fig:XRD_2}. Though it appears that there is a strong $bcc$ presence, a constant spacing of allowed reflections in 2$\theta$ was not observed. These peaks are remnants of the cubic configuration in the sheared monoclinic $\alpha''$ phase. Any identification of the $\gamma$ phase in EBSD patterns is a consequence of the relatively poor patterns close to crystal defects and the pseudo-symmetry problem arising from similar Kikuchi patterns between phases.

\subsection{Microstructural Properties}
Visual inspection using a high resolution optical microscope,  indicated that samples, irrespective of ageing, possessed similar grain sizes, ranging between 50-200\,$\mu$m in diameter. Grains were observed to possess straight edges and have 6 or 7 adjacent grains in the plane. Inclusion particles, primarily niobium carbides \cite{Volz2007}, were found to be randomly dispersed within the material. Through the preparation process, high hardness of these compounds resulted in preferential etching of the metal, causing them to sit proud of the surface.

\begin{figure}
\centering
\includegraphics[width=0.5\textwidth]{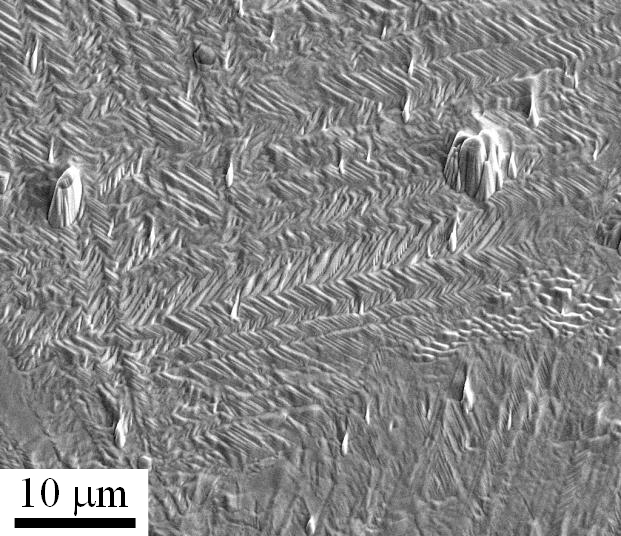}
\caption{Secondary electron image of the surface of the unaged sample after mechanical polishing and electropolishing. Characteristic martensitic herringbone twins are visible. Secondary twins can be observed to run within initial, larger twins.} \label{fig:SEM1}
\end{figure}

\begin{figure}
\centering
\includegraphics[width=0.5\textwidth]{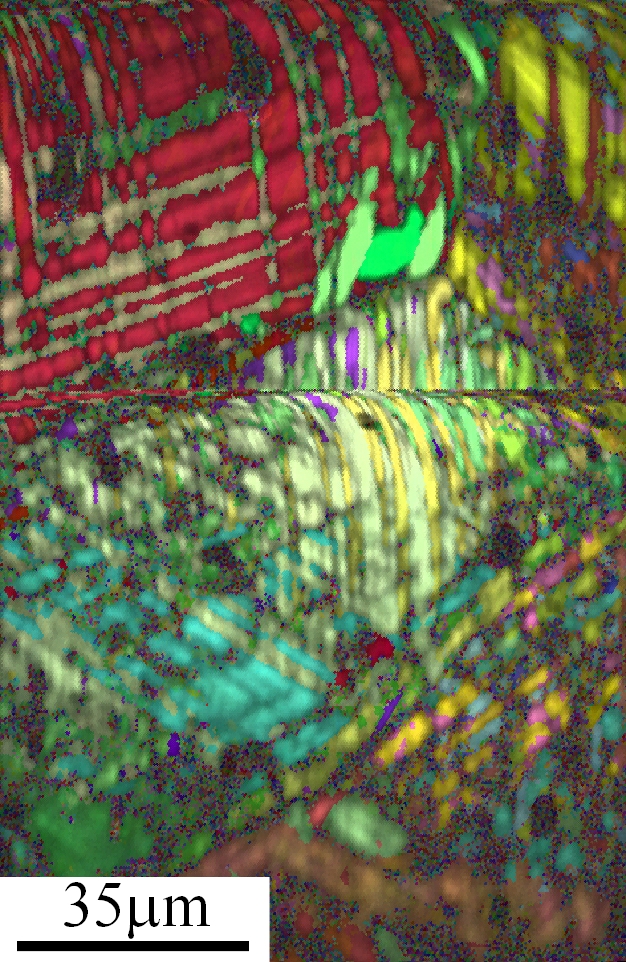}
\caption{EBSD Inverse Pole Figure map of a few grains of the unaged alloy. Levels of twinning is moderate with twins closely spaced together, extending the diameter of grains. Multiple modes have already been induced in the unaged material.} \label{fig:EBSD1}
\end{figure}

\begin{figure}
\centering
\includegraphics[width=0.5\textwidth]{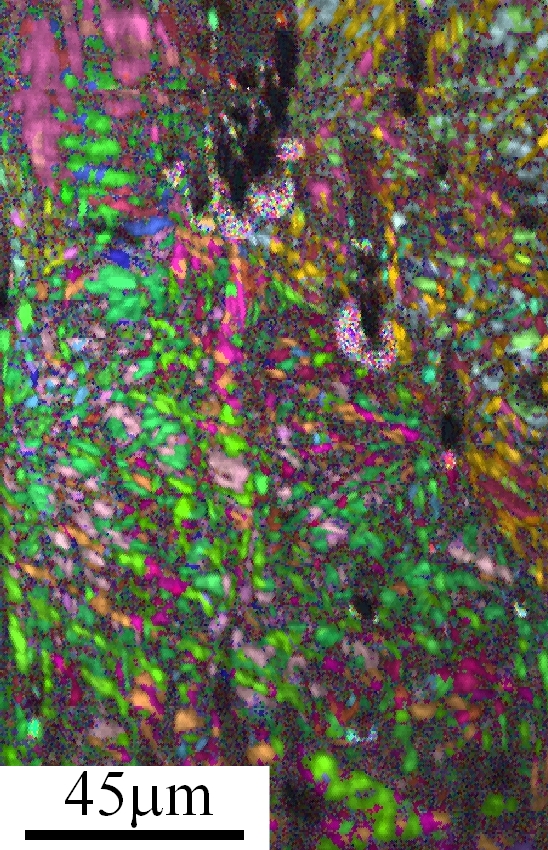}
\caption{EBSD Inverse Pole Figure map of a couple of grains of the 5000 hr aged alloy. Secondary twinning leading to cross cutting and internal distortions are much more prevalent. The mappable fraction has decreased possibly due to the effect of increased preferntial etching from a more complicated microstructure.} \label{fig:EBSD2}
\end{figure}

High magnification secondary electron images of prepared samples revealed fine lineations across the surface of the material, Figure \ref{fig:SEM1}. Etching from electropolishing and/or ion beam steps has preferentially attacked twin boundaries accentuating the herringbone structure known to occur in the martensitic $\alpha''$ \cite{Field2001}. These twins are also readily observed using EBSD, Figures \ref{fig:EBSD1} and \ref{fig:EBSD2}.

Intragranular distortions become increasingly frequent with ageing, as expected \cite{Field2001,Hatt1966}. Further twinning, developing with ageing, resulted in inverse pole figures producing much more vibrant patterns and complicated microstructures within grains, Figure \ref{fig:EBSD2}.

\begin{figure}
\centering
\includegraphics[width=0.5\textwidth]{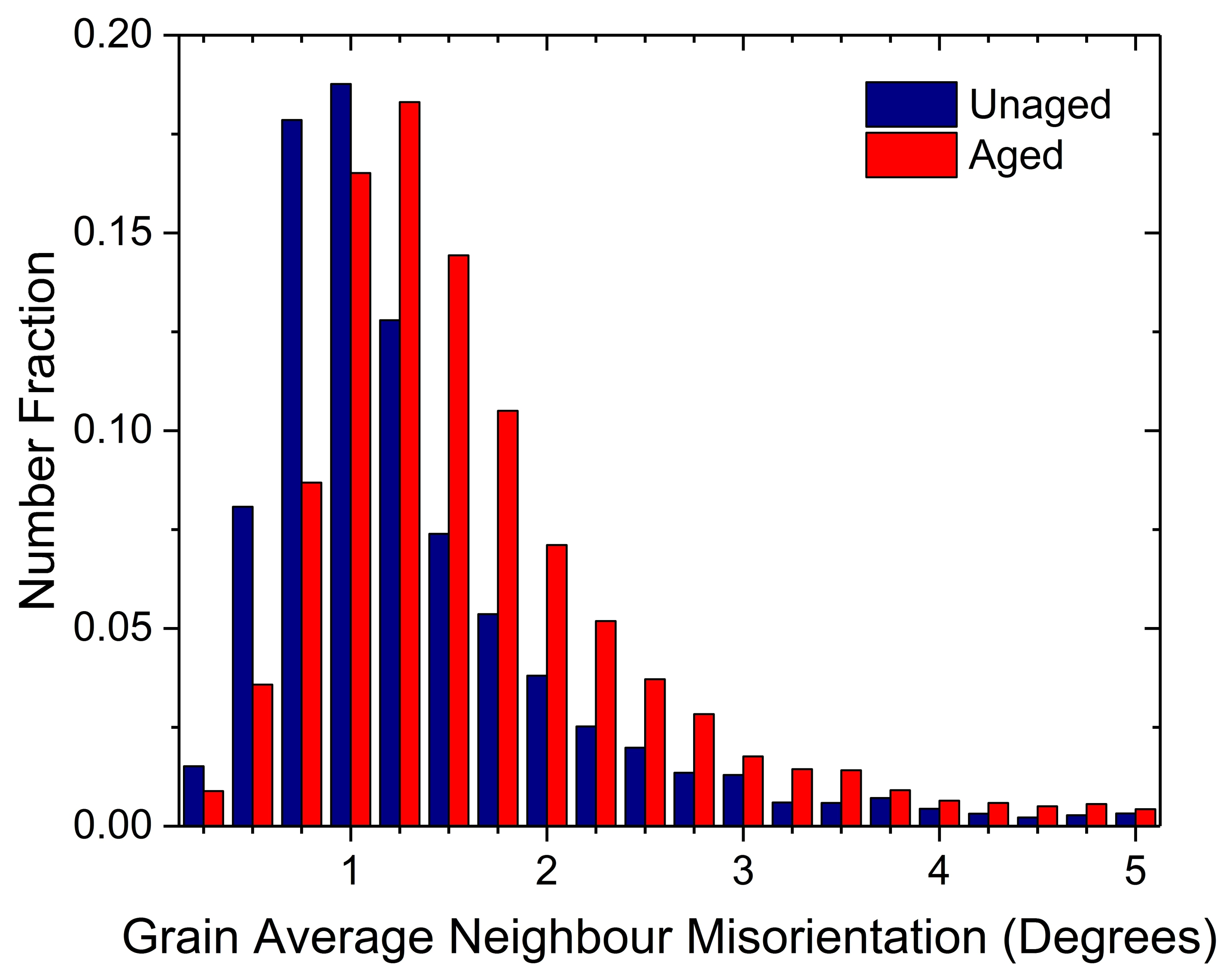}
\caption{GAM plot of the unaged and 3624 hr aged sample. The mean increases by 0.25$^{o}$ between the two ageing periods.} \label{fig:GAM}
\end{figure}

Grain average misorientation (GAM) plots generated for the unaged and 3624\,hr aged samples confirmed crystallographic disturbances within a grain to increase slightly with ageing. Misorientations produced Poisson distributions with the mean increasing with ageing. Distributions for some representative regions of samples are shown in Figure \ref{fig:GAM}. The range was limited from 0$^{o}$ to 5$^{o}$ to remove signals produced from martensitic boundaries \cite{Field2001}. Growing misorientations within martensitic twins highlights the increasingly stressed nature of the material. It is not known whether the measured misorientations are a cause or an effect of increasing martensitic twinning.

\begin{figure}
\centering
\includegraphics[width=0.48\textwidth]{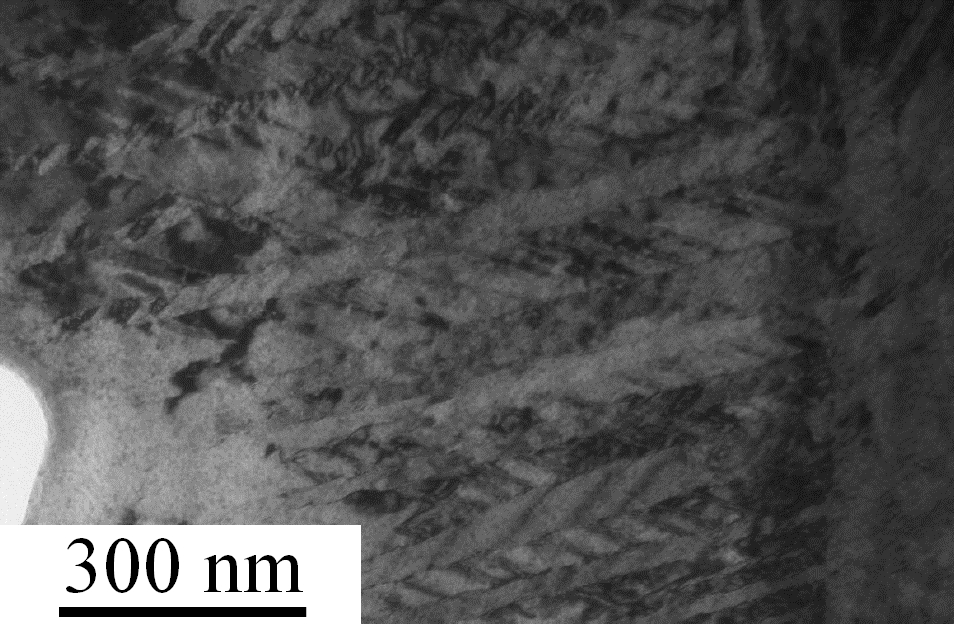}
\caption{Transmission electron micrograph of a UNb5 lamella imaged in BF. Secondary twinning is prevalent with differences in contrast within twins which may be evidence for niobium diffusion driving this process.} \label{fig:TEM1}
\end{figure}

The unaged alloy can easily be recognised as characteristically martensitic in nature with evidence of secondary twinning. Multiple modes have been identified in the literature \cite{Field2001}. Twinning modes such as the \hkl{1-72}, \hkl{-130} and \hkl{021} are possible, creating highly cut up grains, Figure \ref{fig:EBSD1}.

Under relatively low magnification of all samples, TEM lamellae revealed again the twinned microstructure so prevalent in EBSD, Figure \ref{fig:TEM1}. Further evidence complementing SEM was obtained to confirm secondary twinning building in the aged samples. Under high magnification in certain regions, a nanoscale lamination became evident, Figure \ref{fig:TEM3}.

The 5000\,hour aged sample shows the incredibly disruptive effect that ageing has on the structure of the material, Figure \ref{fig:EBSD2}. Further twinning drastically reduces the size of congruent crystalline regions, although analysis of XRD spectra using the Scherrer equation suggests that crystallites are already very small (roughly 20\,nm evaluated for the \hkl(002)$_{\alpha''}$ and \hkl(021)$_{\alpha''}$ peaks) \cite{Patterson1939}.

As with EBSD, TEM suggests that micro- and nanostructures become increasingly complex with extended ageing. Further work will focus on sampling sufficient grains to determine the order and timescales in which these twinning modes evolve.

\subsection{Chemical Composition}
EDX was used in conjunction with SEM to determine the elemental constitution of unaged and 3624\,hr aged sample surfaces. Inclusions were investigated in both samples by EDX analysis and predominately confirmed to be Nb$_{2}$C or UC as seen by Volz et al \cite{Volz2007}. EDX area scans, large enough to transect grain boundaries, were conducted to investigate potential partitioning of niobium to grain boundaries, however no evidence was found to suggest that this was a major chemical redistribution mechanism. Due to the volume of interaction of the beam, errors are larger than any deviation and not capable of providing information regarding small-scale distribution. Despite this, it is possible to say that it is not yet at the stage of significant precipitation of niobium. The formation of oxide, known to form via a two step process, segregating alloy constituents, has also had little effect on the distribution of niobium for the fully prepared samples \cite{Younes2007}.

\begin{figure}
\centering
\includegraphics[width=0.5\textwidth]{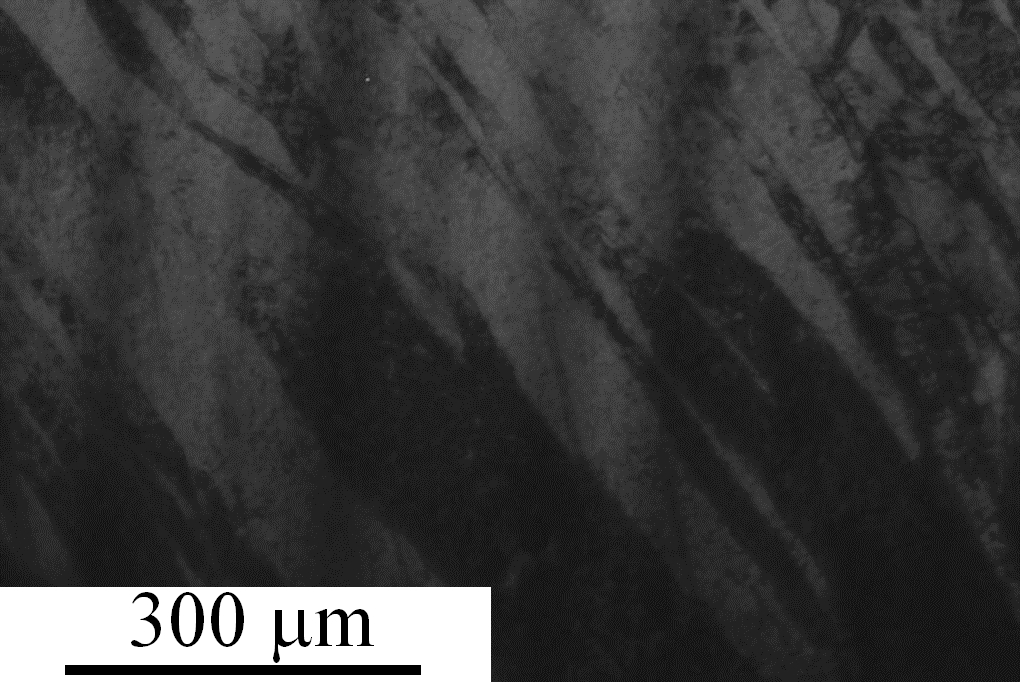}
\caption{Bright field transmission electron micrograph of a UNb5 lamalla. Banding is observed but thought to be too large in scale to be associated with chemical segregation.} \label{fig:TEM3}
\end{figure}

Bright field TEM images, such as that of Figure \ref{fig:TEM3}, showing a 5000\,hr aged sample, may present evidence of banding attributed to chemical segregation of constituent elements, though this would disagree with Clarke et al. \cite{Clarke2009}. Transmittance is greater through regions enriched in niobium creating lighter bands, whilst darker bands correspond to regions deficient in niobium. However, bright field microscopy is also susceptible to channelling contrast produced by varying crystallographic orientations. Given that images such as Figure \ref{fig:TEM1} show structure on these length scales, it is likely that Figure \ref{fig:TEM3} is presenting structural features, but viewed in a non-advantageous direction.

\begin{figure}
\centering
\includegraphics[width=0.5\textwidth]{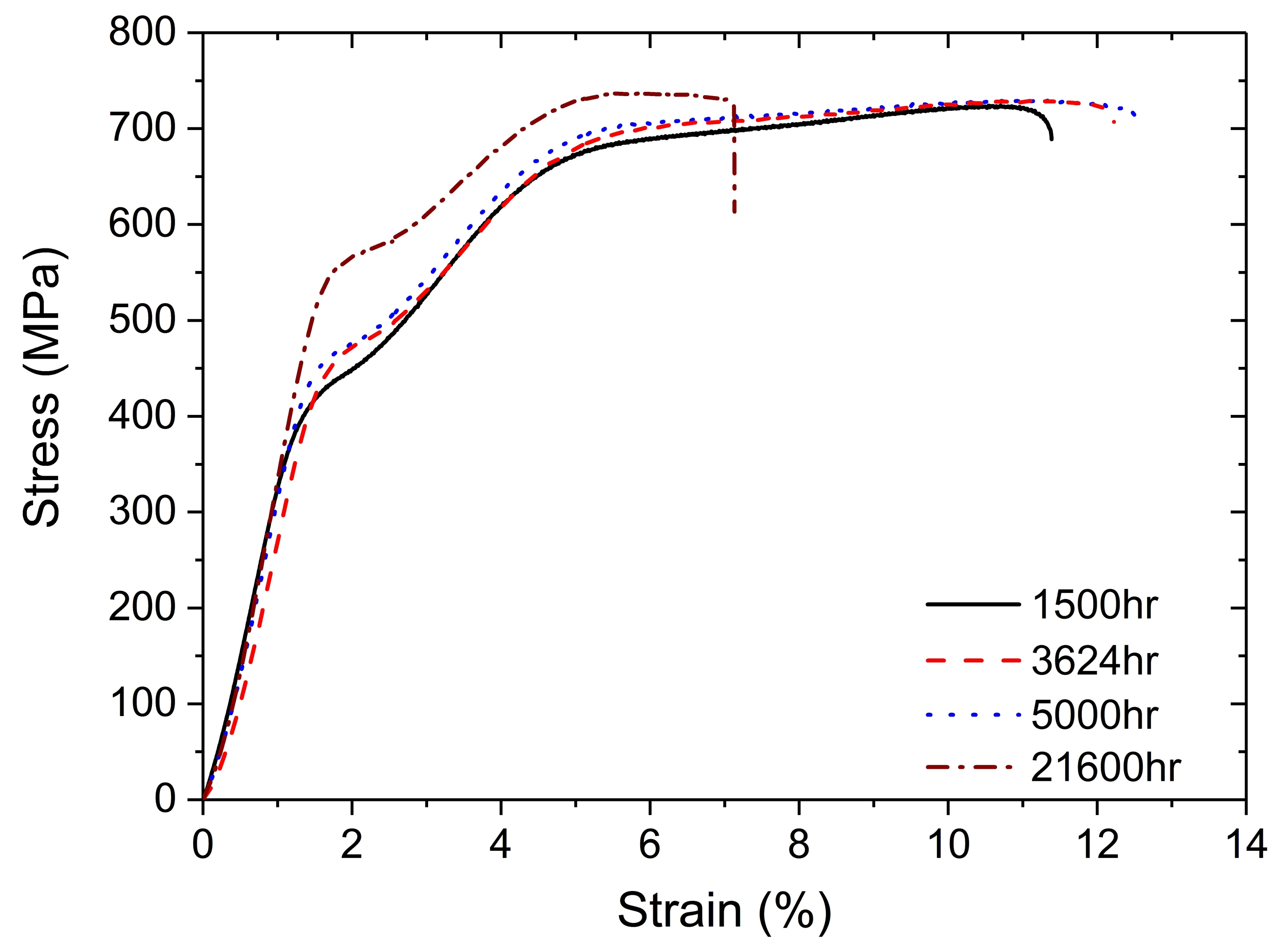}
\caption{Stress-strain curve showing evolution of the UNb5 alloy with an additional measurement of a sample aged at 150\,$\celsius$ for 21600\,hrs.} \label{fig:Stress-strain}
\end{figure}

\subsection{Mechanical Properties}
\begin{figure}
\centering
\includegraphics[width=0.5\textwidth]{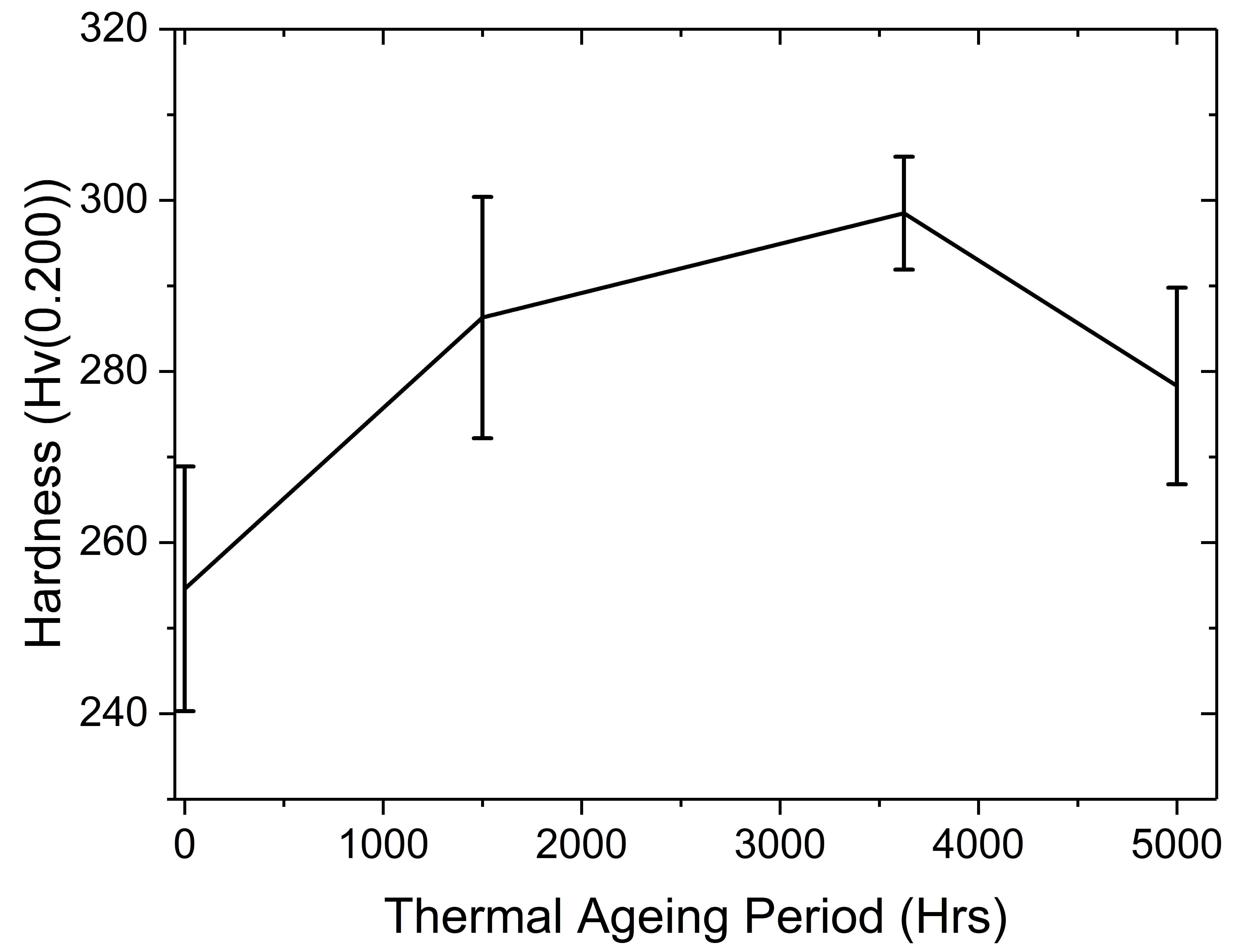}
\caption{Plot showing the assessed relationship between ageing duration and hardness as determined by the Vickers method. Hardness rises slightly between the unaged case and the first ageing period before possibly settling. Large errors are prevalent even after taking numerous measurements, possibly due to the effect of grain orientation.} \label{fig:Hardness}
\end{figure}

Vickers hardness testing showed clear hardening with ageing, Figure \ref{fig:Hardness}, that is consistent with other studies \cite{Volz2007}. The unaged sample produced a hardness of 254.6\,$\pm$\,15.1\,Hv. With ageing, the hardness increased to a maximum of around 300 Hv. Increased twinning in the $\alpha$'' matrix is predicted to have arrested slip type deformations. The rise in hardness points to a slight increase in martensitic characteristics \cite{Nishiyama1978}.

A substantial change to the stress-strain curves is observed between the 5000\,hr and 21600\,hr aged specimens, Figure \ref{fig:Stress-strain}. The 21600\,hr sample experienced a decrease in the fracture point without too much change to the Young's modulus, shape memory properties and upper tensile strength. The properties of the 1500, 3624 and 5000\,hr aged samples are mostly unchanged suggesting that the build up of secondary twins has little effect on the ability for the material to be strained. 

\section{Conclusions}
It is clear that the microstructure of the $\alpha''$ (UNb5) material is inherently complex. With ageing, further twinning within the material can be observed, and is responsible for increased lattice distortion. Martensitic transformations produced in the initial quenching of the material create an incredibly distorted microstructure with induced stresses. Elevated temperatures may have allowed mobility of twins and formation of new ones, relieving stresses.

Microstructural changes are particularly obvious through EBSD and SEM. These may assist in explaining the significantly altered mechanical properties known to, and measured to occur in this material under these conditions. Hardening has been assessed in this work and shown to increase, however, the degree of change is not as severe as in the UNb7 case, which undergoes an isothermal martensitic transformation from an initially `austenitic' state under the same ageing conditions. 

Fundamental mechanisms responsible for primarily microstructural and mechanical changes are thought to have been linked to an increase in martensitic tendencies. The role of diffusion is unclear with TEM, possibly providing evidence of compositional differences at twin boundaries. Further work is required, particularly at longer timescales to understand the substantial changes observed in the stress-strain data of the additional 21600\,hrs aged sample.

\bibliographystyle{unsrt}
\bibliography{U5Paper}

\end{document}